# COSMIC: Emotionally Intelligent Agents to Support Mental and Emotional Well-being in Extreme Isolation: Lessons from Analog Astronaut Training Missions

**Authors:** Dr. A. Xygkou-Tsiamoulou[1], Dr. Alexandra Covaci[2], Zeqi Jia[3], *Prof. Jenny Yiend[3] and *Prof. Chee Siang Ang[1,4,5]

[1] School of Computing, University of Kent, UK
[2] School of Engineering, Mathematics and Physics, University of Kent, UK
[3] Institute of Psychiatry, Psychology and Neuroscience, King's College London, UK
[4] Kent and Medway Medical School, University of Kent, UK
[5] National Heart and Lung Institute, Imperial College London, UK

* J.Y and C.S.A. are co-senior authors

## Abstract

As humanity pivots toward long-duration interplanetary travel, the psychological constraints of Isolated and Confined Environments (ICE) emerge as a primary mission risk. This paper presents **COSMIC** (COmpanion System for Mission Interaction and Communication) representing the inaugural investigation into the deployment of a high-fidelity, emotionally intelligent AI companion in an analog astronaut setting. By integrating a Large Language Model (LLM) architecture with a diffusion-based digital avatar interface, COSMIC transcends traditional task-oriented automation to provide longitudinal affective support. We detail a modular system architecture designed for temporal continuity through short- and long-term memory systems and outline a robust naturalistic observational framework for evaluating psychological resilience at the LunAres Research Station. This work constitutes the first formal submission in the field to evaluate the efficacy of state-of-the-art generative AI and synthesized visual empathy in mitigating the effects of extreme isolation.

## 1. Introduction

The expansion of human presence into deep space necessitates a fundamental reassessment of crew support systems. While historical mission architectures have focused on the physiological and procedural requirements of spaceflight, the psychological toll of prolonged isolation remains a significant and unresolved bottleneck. Crew members operating in Isolated and Confined Environments (ICE) face unique stressors, including disrupted circadian rhythms, sensory monotony, and substantial communication latencies with Earth. These conditions frequently lead to "psychological attrition," characterized by increased stress, loneliness, and a decline in interpersonal cohesion.

This research introduces a pioneering intervention that addresses the critical gap in affective human-AI interaction. To date, robotic and virtual assistants in space, such as CIMON, have primarily functioned as procedural tools or mobile interfaces for technical data. COSMIC (Figure 1 & Figure 2) marks a paradigm shift by establishing the AI as a proactive, emotionally intelligent companion. By leveraging the advanced reasoning capabilities of LLMs and the expressive fidelity of diffusion-generated avatars, the system is designed to simulate a lifelike presence capable of recognizing and responding to complex emotional states. This study represents the first documented effort to deploy such an integrated generative stack in a high-stakes analog mission, providing a blueprint for the future of psychosocial resilience in extreme environments.

**Figure 1**

*COSMIC Avatar (generated by Gemini3)*

**Figure 2**

*Conceptualisation of physically embodied COSMIC (generated by Gemini3)*

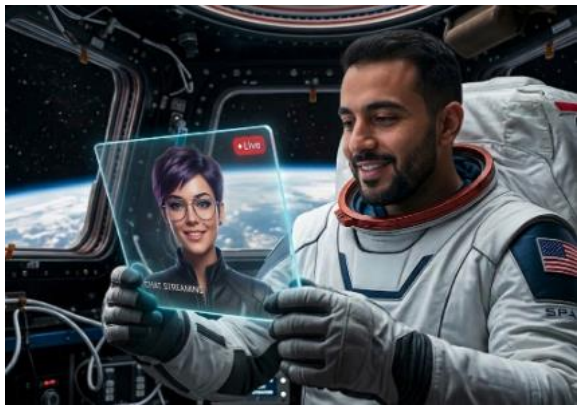

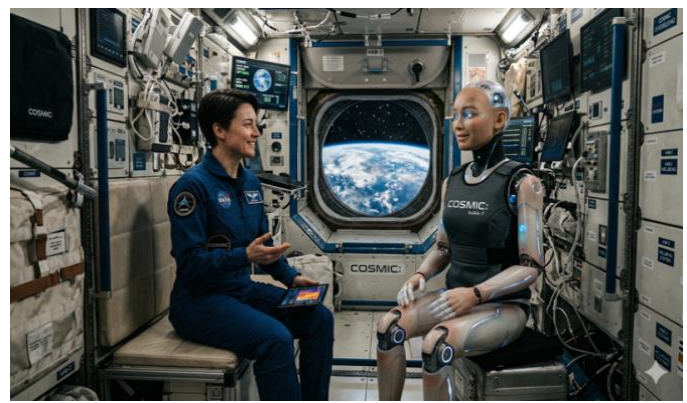

## 2. Conceptual Framework and Related Work

The study of Human-Robot Interaction (HRI) in extreme settings has historically focused on functional autonomy and task efficiency. Early investigations in Antarctic research stations and the International Space Station (ISS) identified the "third-quarter phenomenon," a period of significant morale decline occurring mid-mission. While social robots have been proposed as a remedy, their utility has been limited by the "uncanny valley" (the hypothesis that as an artificial entity becomes more human-like, there is a point where its "almost-human" appearance triggers a sense of unease or revulsion in observers), and the lack of conversational depth inherent in previous generations of Natural Language Processing (NLP).

The emergence of LLMs and diffusion-based generative techniques provides a novel solution to these limitations. Recent literature suggests that high-parameter models can emulate active listening and empathetic dialogue, yet their application in the specific context of astronautics has remained largely theoretical. COSMIC bridges this divide by synthesizing these technologies into a deployed, ecologically valid framework. By moving beyond reactive command-and-control interfaces, this work explores the potential for AI to serve as an "affective buffer," grounding its methodology in the intersection of computational linguistics and social, affective, and cognitive psychology.

## 3. System Architecture and Technical Implementation

The technical foundation of COSMIC is built upon a modular architecture designed to facilitate naturalistic, real-time engagement while maintaining rigorous data logging protocols. At its core, the logic layer utilizes the LLM architecture, which provides the advanced reasoning required for non-clinical, supportive dialogue [4]. This is augmented by a sophisticated memory management system that distinguishes between short-term session coherence and long-term longitudinal recall. This dual-memory approach allows the agent to reference previous mission events and personal disclosures, thereby building a "shared history" with the user; a critical factor in establishing trust and perceived empathy.

The visual layer of COSMIC represents a significant technological advancement in the field. Unlike static or pre-rendered avatars, the diffusion-based interface generates verbal and non-verbal cues in real-time, aligning facial expressions and posture with the emotional tone of the conversation. This multi-modal synthesis ensures that the agent's digital presence is both expressive and contextually appropriate. Furthermore, a dedicated data gateway manages tool invocation, allowing the agent to provide guided meditation, relaxation media, or procedural walkthroughs for Extravehicular Activity (EVA) simulations based on standardized training protocols. This integration ensures that the system is perceived as both an emotional confidant and a functional asset.

## 4. Experimental Design and Methodology

### 4.1 Study Design

To evaluate the impact of COSMIC on astronaut well-being, we employ a comparative, mixed-methods research design. The study utilizes the LunAres COSMIC-enhanced cohort as the experimental group and a historical dataset from the University of Central Florida as the control [3, 5]. This comparative framework allows for the identification of differences in psychological resilience and interaction quality across missions with and without AI-driven affective support.

### 4.2 Participants and Procedure

The experimental phase involves up to 30 analog crew members (n=30) who engage with COSMIC for a minimum of 20 minutes daily over a 14-day mission cycle. Data collection is multi-modal, spanning psychological, physiological, and interactional domains. Psychological metrics are gathered via standardized instruments, including the UCLA Loneliness Scale [1], the Perceived Stress Scale (PSS) [2], the Profile of Mood States (POMS) [7], Human Robot Interaction Evaluation Scale (HRIES) [8], as well as team cohesion scales. These are supplemented by physiological data, such as Heart Rate Variability (HRV) and sleep quality metrics, to provide an objective measure of the stress-buffering effects of the interaction. Post-mission, participants undergo semi-structured interviews to explore the subjective dimensions of trust, social realism, and the perceived empathy of the AI companion.

### 4.3 Ethical Considerations

COSMIC is explicitly designed as a non-clinical intervention; it is not a substitute for professional psychological or psychiatric care but rather a supportive presence for sub-clinical stress and isolation. To protect participant privacy, all conversational logs are pseudonymized at the point of capture, and audio data is deleted immediately following transcription. Furthermore, the agent is programmed to be non-probing and non-intrusive, ensuring that all interactions are participant-led.

## 5. Preliminary Evaluation and Results

Initial testing of the system architecture has demonstrated high levels of user engagement, particularly regarding the responsiveness of the diffusion-based avatar. We hypothesize that the introduction of COSMIC will result in a reduction in perceived loneliness and a stabilization of mood states compared to the historical control group. Furthermore, we expect that the integration of procedural support alongside emotional dialogue will facilitate a unique "trust-building" dynamic. Research indicates that individuals can develop a therapeutic alliance-the collaborative relationship between a helper and a user-even with simple digital applications. We anticipate this effect will be significantly more pronounced with COSMIC, particularly as users begin to associate the agent with a personal identity and name.

### 5.1 Preliminary Data Report

Pilot data from initial mission studies (unpublished) suggest that regular engagement with COSMIC is associated with a consistent reduction in both perceived stress and loneliness (Table 1 & Table 2).

**Table 1**

*Pre- and Post-Mission Psychological Metrics*

| Measure | Pre-Mission Mean | Pre-Mission SD | Post-Mission Mean | Post-Mission SD |
|---|---|---|---|---|
| **PSS (Perceived Stress)** | 18.4 | 3.2 | 17.0 | 2.9 |
| **UCLA Loneliness** | 40.4 | 5.1 | 36.6 | 4.8 |

Daily mood tracking generally supports these patterns. Two participants recorded instances of 'severely' elevated mood early in the mission, potentially reflecting adjustment to the isolation environment. However, crew interaction ratings remained positive overall, with depressed mood, anxiety, and irritability largely reported at mild-to-moderate levels.

Table 2

*User Experience Questionnaire (UEQ) Subscales*

| UEQ subscale | Mean (M) | SD | Definition |
|---|---|---|---|
| Attractiveness | 0.10 | 1.14 | Overall impression of the system: whether it feels pleasant, likeable, and appealing to use. |
| Perspicuity | 1.45 | 0.60 | How easy the system is to understand and learn. |
| Efficiency | 0.15 | 0.65 | How quickly and easily users feel they can get things done with the system. |
| Dependability | 0.35 | 0.72 | How much the system feels reliable, predictable, and under the user's control. |
| Stimulation | -0.20 | 1.19 | How exciting, motivating, and engaging the system feels. |
| Novelty | -0.45 | 0.80 | How innovative, creative, and original the system appears. |

Trust scores showed a cautious but favorable pattern, with low Distrust (M = 2.08) and a moderately positive overall recoded Trust score (M = 3.62). On the social-perceptual scales (HRIES, Table 3), Agency (M = 3.10) and Sociability (M = 2.85) were the strongest dimensions, indicating that COSMIC was perceived as intentional, intelligent, and socially approachable. Lower Animacy (M = 1.95) and relatively low Disturbance (M = 2.30) suggest that while users found the system smart, they did not yet view it as fully "lifelike". Overall, the descriptive findings indicate that COSMIC's main strengths lie in clarity, perceived intelligence, and sociability.

Table 3

*Descriptive statistics for all HRIES measures*

| HRIES subscale | Measure / item | Mean (M) | SD | What it captures |
|---|---|---|---|---|
| **Sociability** | Warm | 2.20 | 0.45 | Perceived warmth and interpersonal friendliness |
| | Likeable | 2.60 | 1.14 | General likeability of COSMIC |
| | Trustworthy | 2.20 | 1.10 | Whether COSMIC seemed worthy of trust |
| | Friendly | 4.40 | 0.89 | Perceived friendliness and social approachability |
| **Sociability summary** | **Subscale mean** | **2.85** | **0.84** | Overall perception of COSMIC as socially pleasant and approachable |
| **Animacy** | Alive | 1.80 | 1.30 | Sense that COSMIC seemed alive |

| HRIES subscale | Measure / item | Mean (M) | SD | What it captures |
|---|---|---|---|---|
| | Natural | 2.20 | 1.10 | Whether COSMIC felt natural rather than artificial |
| | Real | 1.40 | 0.55 | Perceived realism |
| | Human-like | 2.40 | 1.14 | Extent to which COSMIC seemed human-like |
| **Animacy summary** | **Subscale mean** | **1.95** | **1.04** | Overall sense of lifelikeness and naturalness |
| **Agency** | Self-reliant | 2.00 | 0.71 | Whether COSMIC appeared autonomous or self-directed |
| | Rational | 3.80 | 1.79 | Perceived logic and rationality |
| | Intentional | 3.20 | 1.30 | Whether COSMIC seemed to act with intention |
| | Intelligent | 3.40 | 1.52 | Perceived intelligence |
| **Agency summary** | **Subscale mean** | **3.10** | **0.74** | Overall perception of COSMIC as purposeful, intelligent, and agentic |
| **Disturbance** | Creepy | 2.60 | 1.14 | Degree of creepiness or unease |
| | Scary | 1.60 | 0.89 | Extent to which COSMIC felt frightening |
| | Uncanny | 1.80 | 0.84 | Sense of eeriness or uncanny quality |
| | Weird | 3.20 | 1.30 | Perceived strangeness |
| **Disturbance summary** | **Subscale mean** | **2.30** | **0.69** | Overall degree of discomfort or eeriness triggered by COSMIC |

## 6. Discussion

The deployment of COSMIC signals a fundamental transition in the design of space-bound AI systems. By prioritizing affective companionship, the system addresses the "human element" of space exploration that has historically been secondary to technical engineering. The preliminary data positions COSMIC as a promising companion technology with several clear strengths. Most notably, the system was experienced as intelligible and cognitively coherent, suggesting that participants were able to quickly understand how to engage with it as a purposeful conversational agent.

Clarity and perceived rationality provide a strong foundation for sustained interaction. While the quasi-experimental design has limitations-most notably the lack of random assignment-the use of a consistent, high-fidelity analog facility provides a robust environment for establishing the first empirical evidence for generative AI companionship in space. Future evaluations will utilize the Digital Therapeutic Alliance Scale [6] to more accurately measure the depth of the bond formed between the crew and the AI.

## 7. Conclusion

The COSMIC project explores the convergence of state-of-the-art Large Language Models and generative visual avatars in the context of extreme isolation. This work provides a comprehensive framework for the development of future human-AI collaborations, suggesting that emotional intelligence is a critical component for sustaining mission well-being alongside procedural competence.

While the pilot data is preliminary, the findings highlight the importance of developing systems that are not merely functional but socially interpretable. Participants tended to perceive COSMIC as friendly and intentional, indicating that the system is beginning to function as a credible companion. As we look toward the horizon of long-duration spaceflight, the principles established in this study will serve as a foundational guide for maintaining the mental health of those venturing into the furthest reaches of our solar system.

## References

1. Russell, D. W. (1996). UCLA Loneliness Scale (Version 3): Reliability and validity. *Journal of Personality Assessment*.
2. Cohen, S., Kamarck, T., & Mermelstein, R. (1983). A global measure of perceived stress. *Journal of Health and Social Behavior*.
3. NASA. (2025). Human Research Program (HRP) Evidence Reports on Isolated and Confined Environments.
4. OpenAI. (2025). GPT-5 Technical Architecture and Reasoning Capabilities.
5. LunAres Research Station. (2026). Standardized Analog Mission Training Protocols.
6. Tong, F., Lederman, R., D'Alfonso, S., Berry, K., & Bucci, S. (2025). Development of a digital therapeutic alliance scale (MM-DTA) in the context of fully automated mental health apps. *Behaviour & Information Technology*. Advance online publication. https://doi.org/10.1080/0144929X.2025.246967
7. Heuchert, J. P., & McNair, D. M. (2012). *Profile of Mood States 2nd Edition (POMS 2)* Multi-Health Systems.
8. Spatola, N., Kühnlenz, B., & Cheng, G. (2021). Perception and evaluation in human–robot interaction: The Human–Robot Interaction Evaluation Scale (HRIES)—A multicomponent approach of anthropomorphism. *International Journal of Social Robotics*, *13*(7), 1517-1539.

**Correspondence:** Dr. A. Xygkou-Tsiamoulou (A.Xygkou-Tsiamoulou@kent.ac.uk)